\documentclass[reprint,secnumarabic,amssymb,amsmath, nobibnotes,aps,prapplied,superscriptaddress]{revtex4-2}
 
\usepackage{graphicx}
\usepackage{dcolumn}
\usepackage{bm}

\usepackage[utf8]{inputenc}
\usepackage[T1]{fontenc}
\usepackage{mathptmx}
\usepackage{etoolbox}

\usepackage{color,soul} 

\newcommand{\tblue}[1]{{#1}} 

\usepackage{hyperref} 
\hypersetup{colorlinks=true, citecolor=blue, urlcolor=blue, linkcolor=blue}

\begin{document}

\title[]{Multi-functional metasurface architecture for amplitude, polarization and wavefront control}

\author{A. Pitilakis}\email{alexpiti@auth.gr}
\affiliation{ 
School of Electrical and Computer Engineering, Aristotle University of Thessaloniki, GR-54124 Thessaloniki, Greece}%

\author{M. Seckel}
\affiliation{ 
System Integration and Interconnection Technologies, Fraunhofer IZM, 13355 Berlin, Germany}%

\author{A. C. Tasolamprou}
\affiliation{
Institute of Electronic Structure and Laser, Foundation for Research and Technology Hellas, GR-70013, Heraklion, Crete, Greece}%

\author{F. Liu}
\affiliation{Xi'an Jiaotong University, School of Electronic Science and Engineering, 710049, Xi'an, China}%
\affiliation{Department of Electronics and Nanoengineering, Aalto University, FI-00076, Aalto, Finland}%

\author{A. Deltsidis}
\affiliation{
Institute of Electronic Structure and Laser, Foundation for Research and Technology Hellas, GR-70013, Heraklion, Crete, Greece}%

\author{D. Manessis}
\author{A. Ostmann}
\affiliation{ 
System Integration and Interconnection Technologies, Fraunhofer IZM, 13355 Berlin, Germany}%

\author{\\N. V. Kantartzis}
\affiliation{ 
School of Electrical and Computer Engineering, Aristotle University of Thessaloniki, GR-54124 Thessaloniki, Greece}%

\author{C. Liaskos}
\affiliation{Department of Computer Science Engineering, University of Ioannina, Ioannina, Greece}

\author{C. M. Soukoulis}
\affiliation{
Institute of Electronic Structure and Laser, Foundation for Research and Technology Hellas, GR-70013, Heraklion, Crete, Greece}%
\affiliation{Ames Laboratory—U.S. DOE and Department of Physics and Astronomy Iowa State University, Ames, IA 50011, USA}%

\author{S. A. Tretyakov}
\affiliation{Department of Electronics and Nanoengineering, Aalto University, FI-00076, Aalto, Finland}%

\author{M. Kafesaki}
\affiliation{
Institute of Electronic Structure and Laser, Foundation for Research and Technology Hellas, GR-70013, Heraklion, Crete, Greece}%
\affiliation{Department of Materials Science and Technology, University of Crete, GR-70013 Heraklion, Crete, Greece}%

\author{O. Tsilipakos}\email{otsilipakos@iesl.forth.gr}
\affiliation{
Institute of Electronic Structure and Laser, Foundation for Research and Technology Hellas, GR-70013, Heraklion, Crete, Greece}%

\date{\today}

\begin{abstract}
Metasurfaces (MSs) have been utilized to manipulate different properties of electromagnetic waves. By combining local control over the wave amplitude, phase, and polarization into a single tunable structure, a multi-functional and reconfigurable metasurface can be realized, capable of full control over incident radiation. Here, we experimentally validate a multi-functional metasurface architecture for the microwave regime, where in principle variable loads are connected behind the backplane to reconfigurably shape the complex surface impedance. As a proof-of-concept step, we fabricate several metasurface instances with static loads in different configurations (surface mount capacitors and resistors of different values in different connection topologies) to validate the approach and showcase the different achievable functionalities. Specifically, we show perfect absorption for oblique incidence (both polarizations), broadband linear polarization conversion, and beam splitting, demonstrating control over the amplitude, polarization state, and wavefront, respectively. Measurements are performed in the 4-18 GHz range inside an anechoic chamber and show good agreement with theoretically-anticipated results. Our results clearly demonstrate the practical potential of the proposed architecture for reconfigurable electromagnetic wave manipulation.
\end{abstract}

\maketitle

\section{\label{sec:Intro}Introduction}

Metasurfaces, the single-layer version of metamaterials, have attracted considerable interest in recent years \cite{Glybovski:2016,He:2019}. They are capable of manipulating the amplitude, phase, and polarization of the incident electromagnetic wave by appropriately engineering the subwavelength, resonant building blocks (termed meta-atoms). Maximum functionality can be obtained by combining into a single metasurface the ability to locally control all three aforementioned characteristics \cite{Tsilipakos2020progress}. In physical terms, this amounts to locally manipulating the anisotropic, complex surface impedance (real and imaginary parts of the tensor elements)  \cite{Liu:2019}. This approach can lead to multi-functional and reconfigurable metasurfaces, which allow for maximum operation versatility \cite{Pitilakis:2018}. Currently, both static and reconfigurable versions of metasurfaces are being actively researched for a broad range of applications in the microwave and mmWave (5G) frequencies, including absorbers \cite{Assimonis2019}, isolators \cite{Ramaccia2020,Taravati2020}, filters \cite{Hwang2021}, switchable screens \cite{deLustrac2021},  enhanced antennas \cite{Smith2017}, and wavefront shaping devices, which have been both theoretically studied \cite{Yurduseven2020} and experimentally demonstrated \cite{Cai2017,Feng2021}.

In Ref.~\onlinecite{Liu:2019} we conceptualized a metasurface unit cell with locally- and continuously-tunable complex surface impedance, for multiple reconfigurable functions. Further developing this concept, in Ref.~\onlinecite{Pitilakis2021} we proposed a practical and scalable multi-functional metasurface architecture for the microwave regime, where electronic integrated circuits are assembled behind the metasurface backplane in order to dynamically engineer the metasurface properties. The integrated circuits supply tunable resistance and capacitance loads to each meta-atom \cite{Kossifos2020}, allowing to locally shape the complex surface impedance. By co-simulating \cite{Pitilakis2021,Koziel2021} electronic chip and electromagnetic responses, we have theoretically showcased the potential of the proposed structure.

In this work, we report the essential step of experimentally validating the multi-functional metasurface architecture. To this end, we fabricate several instances of the metasurface with static loads mounted behind the backplane in place of the integrated circuits. The fixed surface-mounted devices (SMD), capacitors and resistors, are placed in different connection topologies for each metasurface instance. This allows to  showcase different functionalities of (i)~independent perfect absorption for the two linear  polarizations under oblique incidence, (ii)~polarization control illustrated by broadband linear polarization conversion, and (iii)~wavefront manipulation illustrated by beam splitting. The replacement of the tunable integrated chips with fixed loads serves to experimentally validate the metasurface architecture prior to the costly assembling of the actual chips. Note that the chosen values of resistance and capacitance are within the capabilities of the custom chip implementation  \cite{Pitilakis2021,Kossifos2020}, so that the  demonstrated functionalities are readily achievable with the reconfigurable version of the proposed multi-functional metasurface.  

The remainder of the paper is organized as follows. In Sect.~\ref{sec:Architecture} we present the metasurface architecture and discuss fabrication and measurement details. Simulation and experimental results are presented in Sect.~\ref{sec:Results} for all three showcased functionalities. Finally, the conclusions appear in Sect.~\ref{sec:Conclusions}.

\section{\label{sec:Architecture}Multi-functional metasurface architecture}

\begin{figure}
	\centering
	\includegraphics{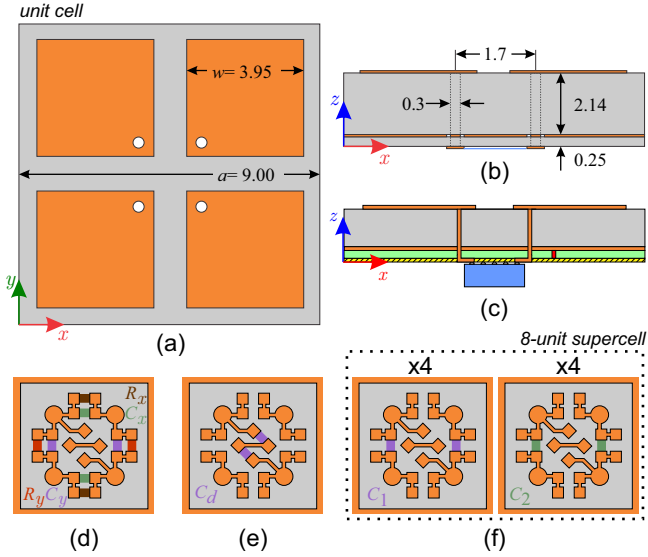}
	\caption{Architecture of the proposed reconfigurable metasurface: (a)~Top view: unit cell with four patches and (b)~side view: three-metallization-layer stackup (dimensions in mm). (c)~Reconfigurable version with chip. Static load configuration in the bottom metallization layer for (d)~independent perfect absorption for the two linear polarizations: different resistor-capacitor, $RC$, pairs connect the adjacent patches in the unit cell along $x$ and $y$ directions; (e)~linear polarization conversion: appropriate capacitance connects the diagonally-adjacent patches in the unit cell;  (f)~beam splitting: supercell of unit cells with different capacitances. \label{fig:Architecture}}
\end{figure}

The multi-functional metasurface under study is a three-metallization-layer printed circuit board (PCB) structure based on a high-frequency Panasonic Megtron7N dielectric substrate ($\varepsilon_r=3.35$, $\tan\delta=0.002$); the geometry is described in Fig.~\ref{fig:Architecture}. The top metallization layer contains a $2\times2$ array of square copper patches ($w\times w=3.95$~mm~$\times$~$3.95$~mm) in a symmetric configuration [Fig.~\ref{fig:Architecture}(a)]; the periodicity is square with lattice constant  $a=9$~mm ($\sim\lambda_0/7$ at the frequency of 5~GHz). The second (middle) metallization layer is the metasurface backplane so that the structure operates in reflection (negligible transmission). Finally, the third (bottom) metallization layer accommodates the SMD components [Fig.~\ref{fig:Architecture}(b,d-f)], or the integrated chip in the reconfigurable version [Fig.~\ref{fig:Architecture}(c)]. This way, the loads are ``hidden'' beneath the backplane and do not interfere with the incident electromagnetic waves nor obstruct the aperture; connection between patches and loads is accomplished by means of through vias (TVs). The fabricated \cite{Manessis2020,Manessis2021} metasurface boards consisted of $18\times26$ cells, for an effective aperture of 162~mm $\times$ 234~mm. Photographs of both sides of the metasurface boards can be found in the Supplemental Material~\footnote{See the Supplemental Material for (i)~conceptual strategy for multi-functionality, (ii)~implementation details for full-wave simulation at the unit-cell level, (iii)~additional photos of fabricated samples and measurement setup, (iv)~simplified model and underlying resonances of polarization conversion functionality, (v)~cross-polarization response for oblique incidence, (vi)~unit-cell states for beam splitting functionality, (vii)~details of Huygens-Fresnel methodology for obtaining the farfield radiation pattern.} [Fig.~S4(a-b)].

The four-patch unit cell geometry along with the possibilities for horizontal, vertical, and diagonal connections behind the backplane has been judiciously chosen so as to allow for electrically bestowing anisotropic properties in the surface impedance. Specifically, it provides the ability to address the $x$- and $y$-linear polarization independently (e.g. absorb at different frequencies or incident angles), as well as  allow for linear polarization conversion. The current patterning of the bottom copper layer can accommodate two components in parallel for each of the $x$- and $y$-directions, and one series connection in the diagonal branch [Fig.~\ref{fig:Architecture}(d-f)]. 

These possibilities offered by our metasurface architecture allow for a broad range of functionalities. Populating the vertical and horizontal SMD slots with appropriate combinations of resistance and capacitance pairs ($R_y,C_y$ and $R_x,C_x$, respectively), as shown in Fig.~\ref{fig:Architecture}(d), we can selectively absorb the two linear polarizations for different frequencies and/or incidence angles. Populating the diagonal branch [Fig.~\ref{fig:Architecture}(e)], we can couple the two orthogonal linear polarizations and achieve polarization conversion; by an appropriate capacitance $C_\mathrm{tot}=C_d/2$ we can tune the supported resonances and achieve large aggregate spectral bandwidths. Finally, we can form supercells by assembling different components in the constituent unit cells. This is exemplified in Fig.~\ref{fig:Architecture}(f) where a supercell comprised of 8 9-mm-wide unit cells is formed from two blocks (4 unit cells each) with different capacitance value in the vertical connections ($C_1$ and $C_2$, respectively). The supercell dimension exceeds the free space wavelength allowing to split the output beam, equivalent with a binary grating structure. For more details regarding our generic strategy towards multiple functionalities see Supplemental Material, Fig.~S1.


\begin{figure}
	\centering
	\includegraphics{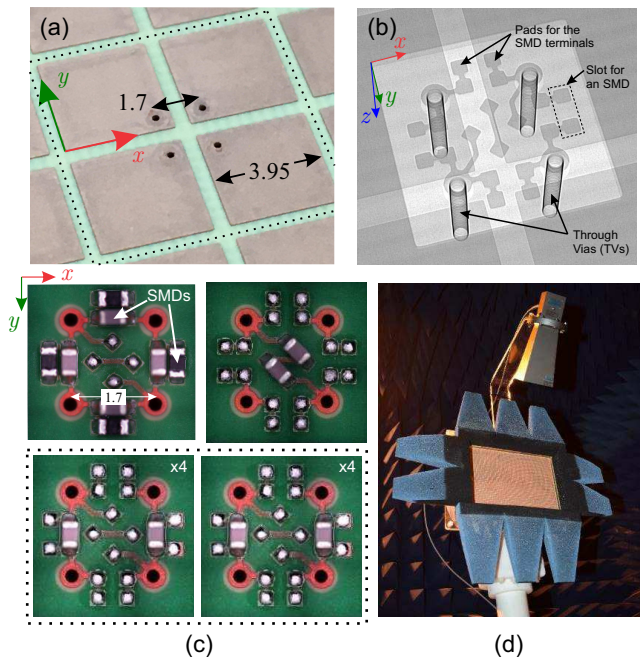}
	\caption{Fabricated samples and measurement setup. (a)~Photograph of fabricated sample (top side). (b)~X-ray view verifying electrical continuity between patches, through vias and SMD pads. (c)~SMD components assembled in the back side for the three different scenarios schematically depicted in Fig.~1(d-f). 
	Top left: $RC_y=22~\Omega\parallel 2.7~\mathrm{pF}$ and $RC_x=100~\Omega\parallel 0.8~\mathrm{pF}$.
	Top right: $C_d=0.8~\mathrm{pF}$. Bottom: $C_1=0.8~\mathrm{pF}$ and $C_2=2.7~\mathrm{pF}$. (d)~Measurement setup inside the anechoic chamber. Metasurface mounted on the positioner head along with a horn antenna at a fixed angle.  \label{fig:FabAndMeas}}
\end{figure}

The fabricated samples are depicted in Fig.~\ref{fig:FabAndMeas}. They concern  three different instances of the metasurface with different SMD components assembled on the back side [Fig.~\ref{fig:FabAndMeas}(c)] according to the three scenarios depicted in Fig.~\ref{fig:Architecture}(d-f). For the specific component values in each case see the caption of Fig.~\ref{fig:FabAndMeas}. A top view of the fabricated samples focusing on the 4-patch unit cell is depicted in Fig.~\ref{fig:FabAndMeas}(a). A corresponding X-ray view is depicted in Fig.~\ref{fig:FabAndMeas}(b); this inspection was used to verify the electrical continuity between patches, TVs and SMD pads. 
The metasurface mounted for measurement inside the anechoic chamber is depicted in Fig.~\ref{fig:FabAndMeas}(d). The assembled boards are mounted on the head of a motorized positioner, allowing for rotation of its mast and head. Standard-gain pyramidal horns are used as transmitting (Tx) and receiving (Rx) antennas; they are mounted either on tripods or on an arm attached on the rotating positioner mast or head [as is the case in Fig.~\ref{fig:FabAndMeas}(d)], for obtaining 2D or 3D scattering pattern measurements. A vector network analyzer (Anritsu 37397D) feeds the horn antennas to perform S-parameter farfield measurements of the metasurface and of a reference reflective plate (of equal dimensions), used for normalization. A photograph of the whole bistatic measurement setup inside the anechoic chamber can be found in the Supplemental Material [Fig.~S4(c)].




\section{Experimental demonstration of multiple functionalities\label{sec:Results}}

\subsection{Polarization- and Direction-sensitive Absorber}

The first metasurface functionality studied is perfect absorption, where the uniform $RC$ loadings determine the resonance frequency and  resonance ``depth'', primarily governed by the capacitance ($C$) and resistance ($R$), respectively. Appropriate $RC$ pairs (SMD components) for two distinct resonances near 5~GHz were assembled on the two parallel $x$- and $y$-slots of each board, as illustrated in Fig.~\ref{fig:Architecture}(d) and Fig.~\ref{fig:FabAndMeas}(c) [top left panel]. We measure the co-polarized reflection spectrum $r=S_{21}$, where a minimum in reflection corresponds to an absorption maximum for our uniform metal-backed metasurface (no transmission and no diffraction orders allowed); note that  cross-polarization coupling is negligible for this configuration, confirmed by both simulation and experiment. 

\begin{figure}
	\centering
	\includegraphics[width=8cm]{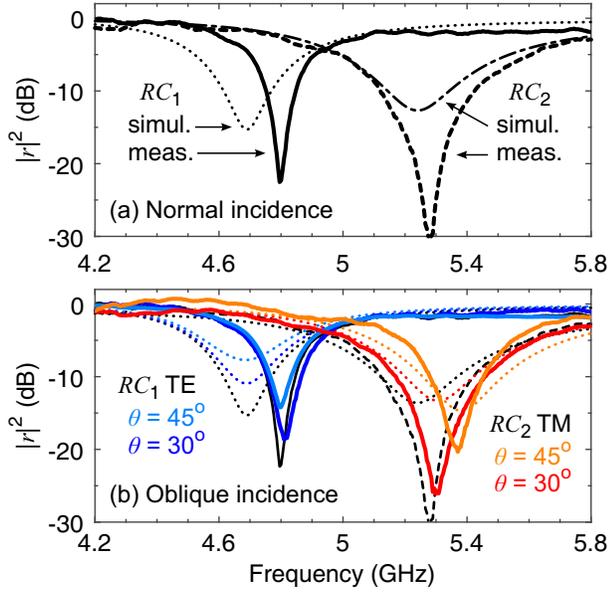}
	\caption{Comparison of simulated and measured reflection spectra for the perfect absorption functionality. The $RC$ loads in vertical and horizontal slots are {$RC_1=22~\Omega\parallel2.7$~pF and $RC_2=100~\Omega\parallel0.8$~pF}, leading to resonances near 4.8 and 5.3~GHz when the electric field is polarized along the $y$ and $x$ axis, respectively. (a) Normal incidence. The resonance frequencies are nicely reproduced in the experiment; the deeper reflection minima in the measurements are due to additional loss compared to the simulation (see text). (b)~Oblique incidence inside the $xz$ plane: TE polarization when $E\parallel y$ involving the $RC_1$ loads and TM polarization when $E\parallel x$ involving the $RC_2$ loads. As the incidence angle increases, the TE resonance depth decreases and the TM resonance frequency is blue shifted. \label{fig:Abs}}
\end{figure}

Figure~\ref{fig:Abs}(a) depicts results for normal incidence; simulation spectra were obtained through single-cell simulations (periodic boundary conditions) performed in CST Studio; specific details can be found in the Supplemental Material, Section~S3. When the electric field is $y$-polarized($x$-polarized), it is the $RC_1$($RC_2$) pairs in the vertical(horizontal) slots that govern the resonance. The measurement verifies that the resonance frequencies are to the left and right of 5~GHz in the two cases, as designed; small blueshifts in resonance frequency are attributed to inductive reactance from soldering of the SMD elements, more pronounced in the case of the 2.7~pF capacitor ($RC_1$). Moreover, the measured reflection dips are deeper than those predicted in the simulations, which is attributed to extra losses stemming from parasitic resistance of the SMD capacitors, soldering, and PCB materials. In anticipation of such additional resistive contributions, the nominal resistance values were chosen such that the metasurface is in the ``undercoupled'' regime, meaning  that any extra resistance would push the operation point towards critical coupling and deepen the reflection minimum  \cite{Zhang2020,Christopoulos2019}. 

Results for oblique incidence inside the $xz$ plane are depicted in Fig.~\ref{fig:Abs}(b). In the case denoted by $RC_1$-TE($RC_2$-TM), the electric field is polarized along the $y$ axis($x$ axis) leading to TE(TM) polarization. In all cases, the receiver horn is aligned at the specular reflection direction, since the metasurface is uniform and no higher diffraction orders are propagating. In the simulations, we observe that for the TE oblique incidence, as the angle increases the  resonance depth diminishes without a visible frequency shift; the opposite holds for the TM-polarized case: the resonance frequency is visibly shifted (increased) without a change in the resonance depth. These trends are reproduced in the experiment corroborating the reasonably good agreement between simulation and measurement. \tblue{A small discrepancy is observed for the TM polarization where the measured reflection minima become shallower as the incidence angle increases. However, this apparent change concerns small values: from $-30$~dB to $-21$~dB. Note that some discrepancy between simulation and measurement is expected since simulations refer to the infinitely-periodic metasurface, whereas measurements are conducted with a finite-size metasurface: $18 \times 26$ cells (162~mm $\times$ 234~mm). This discrepancy will be more pronounced for large incidence angles, since a smaller effective aperture is captured in this case.}


\subsection{Broadband Polarization Converter}

The second functionality studied is broadband linear-polarization conversion in reflection by electrically rotating the principal axis of the surface impedance of the metasurface. The metasurface is again loaded uniformly, but the loading is placed diagonally inside the square unit cell, as in Fig.~\ref{fig:Architecture}(e) and the top right panel of Fig.~\ref{fig:FabAndMeas}(c), so as to emulate a $45^\circ$-cut wire \cite{Grady:2013} that couples orthogonal ($x$ and $y$) linear polarizations. We have also assessed the combination of populating one horizontal and one vertical connection in the back of the cell, emulating an ``L'' shape geometry which has been also successfully used for polarization conversion \cite{Levesque2014}, but found inferior performance. This time we only use capacitor loading, i.e., no resistors, to minimize absorption. Following a parametric simulation study, we found that the capacitance value required for broadband and high-amplitude cross-polarized reflection was around 0.5~pF. Since the minimum $C$-value of available commercial SMD components was in the order of 1~pF, we combine two SMD capacitors in series to attain the required value. 

\begin{figure}
	\centering
	\includegraphics[width=8cm]{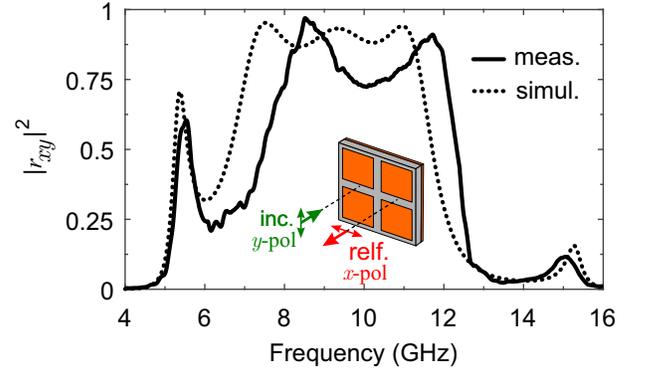}
	\caption{Simulated and measured cross-polarized reflection spectra under normal incidence for the metasurface with diagonal loading consisting of two 0.8~pF capacitors in series. Note the broad high-amplitude cross-polarized spectrum covering the entire X-band (8-12~GHz) and the two isolated peaks near 5.5 and 15~GHz, which are also reproduced in the experiment.	\label{fig:Pol}}
\end{figure}

Figure~\ref{fig:Pol} presents the simulated and measured cross-polarized reflection (XPR) spectra, under normal incidence. The broadband XPR covering the entire X-band (8-12 GHz) arises as a combination of distinct features merging in a continuous aggregate band \cite{Tsilipakos2021}. In the Supplemental Material {(Fig.~S3)} we investigate the field profiles of the individual resonances adopting a simplified metasurface model. \tblue{The width of the main cross-polarized reflection band is found in good agreement between measurement and simulations and approaches $\sim 5$~GHz. However, the spectral features within the XPR band are not completely matching. We attribute this discrepancy to the utilization of several sets of horn antennas (four in total) for covering the frequency ranges 4-6, 6-8, 8-12, 12-18 GHz (approximately), in order to perform this wideband measurement. The four measurements were subsequently stitched together to end up with the result depicted in Fig.~\ref{fig:Pol}. In particular, peaks near the edges of each measurement may be somewhat distorted. We think that this is the case with the first and second peaks, which are close to the stitching (horn swapping) at 8.2 GHz. This hypothesis is corroborated by the fact that the outlying peaks near 5.5 and 15~GHz show excellent agreement between simulation and measurement.}
Finally, we note that for oblique incidence the bandwidth of XPR deteriorates; for angles above $15^\circ$, two reflection dips slice the XPR spectrum in three bands, as can be seen in the Supplemental Material {(Fig.~S5)}.

\subsection{Wavefront Shaping}

The last functionality studied is wavefront manipulation, exemplified through the case of beam splitting: A normally incident beam (plane wave) is divided in two approximately equal beams in symmetric oblique directions, i.e., $\pm\theta_s$. This is accomplished by applying a (non-uniform) binary encoding across the aperture, as shown in Fig.~\ref{fig:Architecture}(f) and bottom panel of Fig.~\ref{fig:FabAndMeas}(c). Specifically, two different capacitor loads are identified which lead to reflection coefficients that exhibit a near-unity amplitude and a $180^\circ$ phase difference at the operating frequency; for details, see Supplemental Material, Section~S6. For  our unit cell design and target band of 4-6 GHz, the two required capacitance values were in the vicinity of 1 and 3~pF, respectively. Subsequently, we assemble these loads so as to form supercells whose extent is larger than the wavelength ($\lambda<p<2\lambda$, $p$ is the supercell period) and implement a flat binary grating on the metasurface (a ``stripes'' pattern), which leads to first-order diffraction modes (scattered beams) in directions $\theta_s=\pm\mathrm{sin}^{-1}\{\lambda/p\}$, assuming normal incidence and infinite aperture. In our case, 
we opted for a period of $p=8a=72$~mm made from eight 9-mm-wide unit cells arranged in four columns of cells with $C_1=0.8$~pF loading followed by four columns of cells with $C_2=2.7$~pF loading; each cell contained two identical SMD capacitors placed in the vertical slots as shown in Fig.~\ref{fig:Architecture}(f) and the bottom panel of Fig.~\ref{fig:FabAndMeas}(c). When the illuminating field is polarized parallel to these loads, the infinite grating produces diffraction lobes approximately in the $\theta_s=\pm51^\circ$ directions; longer periods lead to diffraction closer to specular reflection (normal, in our case), but could not be well-accommodated inside our finite metasurface. 

\begin{figure}
	\centering
	\includegraphics[width=8cm]{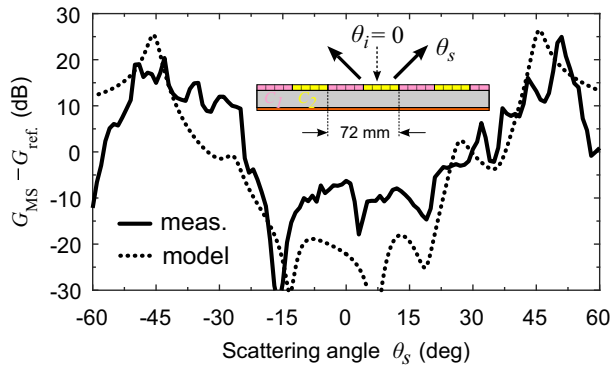}
	\caption{Semi-analytically predicted and measured co-polarized scattering \tblue{gain} pattern at 5.3~GHz for a normally illuminated metasurface configured as a binary grating. The grating is inscribed using two reactive loadings (0.8 and 2.7~pF) that exhibit a $\pi$ phase-difference at the specified frequency. The supercell period is $p=8a=72$~mm which results in a two-beam splitting at approximately $\theta_s=\pm51^\circ$ ($\pm1$ diffraction orders). \label{fig:Split}}
\end{figure}

The measured and semi-analytically extracted 2D scattering patterns, depicted as the metasurface \tblue{gain} normalized to the reference reflector (of same aperture), are illustrated in Fig.~\ref{fig:Split}, \tblue{showing overall good agreement, despite some degradation near the splitting maxima}. In this case, the theoretical results are not from full-wave simulations but are based on the Huygens-Fresnel principle \cite{Taghvaee2020} that estimates the scattered farfield pattern from the unit cell reflection coefficients using Fraunhofer diffraction superposition; implementation details \tblue{for modeling the metasurface and the absorbing-foam frame seen in Fig.~\ref{fig:FabAndMeas}(d)} can be found in the Supplemental Material {(Section~S7)}. The asymmetry in the scattering pattern with respect to the ``right'' and ``left'' lobe is due to the incommensurate number of cells between the metasurface aperture (26) and the grating period (8) in combination with the partially reflecting material that frames the effective aperture [Fig.~\ref{fig:FabAndMeas}(d)]; more details can be found in the Supplemental Material. This is also the cause of the split-lobe maxima appearing at slightly different angles than the prescribed ($\theta_s=\pm51^\circ$) in simulation and measurement.


Finally, we note that the very same structure can be used for beam steering, or anomalous reflection, as outlined in the Supplemental Material {(Fig.~S9)}.

\section{Conclusions\label{sec:Conclusions}}
In conclusion, with this family of static-load metasurfaces we experimentally verify our approach towards a microwave multi-functional and reconfigurable metasurface. The simulation and experimental results have demonstrated successful amplitude, wavefront and polarization control. More generally, this study serves as a proof-of-concept for the broader, software-controlled intelligent metasurface vision, i.e., when the resistive \emph{and} capacitive loads are supplied by chips embedded in the unit cells, forming an inter-connected controller network which is computer controlled \cite{Tsilipakos2020progress}. \tblue{The measured performance using commercial-off-the-shelf (COTS) SMD loads exceeded our expectations, particularly for the demanding absorber functionality; consequently, we anticipate similar or improved in-band performance when using custom-designed chips instead of COTS components. Note that embedding such chips in the back-side of the unit-cells \cite{Pitilakis2021,Kossifos2020} will not perturb the EM design and performance of the metasurface, owing to the decoupling offered by the backplane, nor obstruct its aperture.}


\section*{acknowledgments}
Support by the European Union Horizon 2020 Research and Innovation Programme-Future Emerging Topics (FET Open) under Grants 736876 (project VISORSURF) and 829061 (project NANOPOLY). 
Support by the European Union and Greek national funds through the Operational Program Competitiveness, Entrepreneurship and Innovation, under the call ``Research--Create--Innovate'' (Project code: No. T1EDK-02784). 
Support by the Hellenic Foundation for Research and Innovation (H.F.R.I.) under the ``2nd Call for H.F.R.I. Research Projects to support Post-doctoral Researchers'' (Project Number: 916, PHOTOSURF). 
The authors acknowledge Prof. Julius Georgiou for insights on compatibility of cell designs with custom-made controllable chips as cell loads, Dr. Georgios Pyrialakos and Dr. Michail Christodoulou, for anechoic chamber setup, and the entire VISORSURF project Consortium for useful discussions.






\bibliography{PolBoards}

\begin{thebibliography}{26}%
\makeatletter
\providecommand \@ifxundefined [1]{%
 \@ifx{#1\undefined}
}%
\providecommand \@ifnum [1]{%
 \ifnum #1\expandafter \@firstoftwo
 \else \expandafter \@secondoftwo
 \fi
}%
\providecommand \@ifx [1]{%
 \ifx #1\expandafter \@firstoftwo
 \else \expandafter \@secondoftwo
 \fi
}%
\providecommand \natexlab [1]{#1}%
\providecommand \enquote  [1]{``#1''}%
\providecommand \bibnamefont  [1]{#1}%
\providecommand \bibfnamefont [1]{#1}%
\providecommand \citenamefont [1]{#1}%
\providecommand \href@noop [0]{\@secondoftwo}%
\providecommand \href [0]{\begingroup \@sanitize@url \@href}%
\providecommand \@href[1]{\@@startlink{#1}\@@href}%
\providecommand \@@href[1]{\endgroup#1\@@endlink}%
\providecommand \@sanitize@url [0]{\catcode `\\12\catcode `\$12\catcode
  `\&12\catcode `\#12\catcode `\^12\catcode `\_12\catcode `\%12\relax}%
\providecommand \@@startlink[1]{}%
\providecommand \@@endlink[0]{}%
\providecommand \url  [0]{\begingroup\@sanitize@url \@url }%
\providecommand \@url [1]{\endgroup\@href {#1}{\urlprefix }}%
\providecommand \urlprefix  [0]{URL }%
\providecommand \Eprint [0]{\href }%
\providecommand \doibase [0]{https://doi.org/}%
\providecommand \selectlanguage [0]{\@gobble}%
\providecommand \bibinfo  [0]{\@secondoftwo}%
\providecommand \bibfield  [0]{\@secondoftwo}%
\providecommand \translation [1]{[#1]}%
\providecommand \BibitemOpen [0]{}%
\providecommand \bibitemStop [0]{}%
\providecommand \bibitemNoStop [0]{.\EOS\space}%
\providecommand \EOS [0]{\spacefactor3000\relax}%
\providecommand \BibitemShut  [1]{\csname bibitem#1\endcsname}%
\let\auto@bib@innerbib\@empty
\bibitem [{\citenamefont {Glybovski}\ \emph {et~al.}(2016)\citenamefont
  {Glybovski}, \citenamefont {Tretyakov}, \citenamefont {Belov}, \citenamefont
  {Kivshar},\ and\ \citenamefont {Simovski}}]{Glybovski:2016}%
  \BibitemOpen
  \bibfield  {author} {\bibinfo {author} {\bibfnamefont {S.~B.}\ \bibnamefont
  {Glybovski}}, \bibinfo {author} {\bibfnamefont {S.~A.}\ \bibnamefont
  {Tretyakov}}, \bibinfo {author} {\bibfnamefont {P.~A.}\ \bibnamefont
  {Belov}}, \bibinfo {author} {\bibfnamefont {Y.~S.}\ \bibnamefont {Kivshar}},\
  and\ \bibinfo {author} {\bibfnamefont {C.~R.}\ \bibnamefont {Simovski}},\
  }\bibfield  {title} {\bibinfo {title} {Metasurfaces: From microwaves to
  visible},\ }\href {https://doi.org/10.1016/j.physrep.2016.04.004} {\bibfield
  {journal} {\bibinfo  {journal} {Phys. Rep.}\ }\textbf {\bibinfo {volume}
  {634}},\ \bibinfo {pages} {1} (\bibinfo {year} {2016})}\BibitemShut {NoStop}%
\bibitem [{\citenamefont {He}\ \emph {et~al.}(2019)\citenamefont {He},
  \citenamefont {Sun},\ and\ \citenamefont {Zhou}}]{He:2019}%
  \BibitemOpen
  \bibfield  {author} {\bibinfo {author} {\bibfnamefont {Q.}~\bibnamefont
  {He}}, \bibinfo {author} {\bibfnamefont {S.}~\bibnamefont {Sun}},\ and\
  \bibinfo {author} {\bibfnamefont {L.}~\bibnamefont {Zhou}},\ }\bibfield
  {title} {\bibinfo {title} {Tunable/reconfigurable metasurfaces: Physics and
  applications},\ }\href@noop {} {\bibfield  {journal} {\bibinfo  {journal}
  {Research}\ }\textbf {\bibinfo {volume} {2019}},\ \bibinfo {pages} {1849272}
  (\bibinfo {year} {2019})}\BibitemShut {NoStop}%
\bibitem [{\citenamefont {Tsilipakos}\ \emph {et~al.}(2020)\citenamefont
  {Tsilipakos}, \citenamefont {Tasolamprou}, \citenamefont {Pitilakis},
  \citenamefont {Liu}, \citenamefont {Wang}, \citenamefont {Mirmoosa},
  \citenamefont {Tzarouchis}, \citenamefont {Abadal}, \citenamefont {Taghvaee},
  \citenamefont {Liaskos}, \citenamefont {Tsioliaridou}, \citenamefont
  {Georgiou}, \citenamefont {Cabellos-Aparicio}, \citenamefont {Alarc{\'{o}}n},
  \citenamefont {Ioannidis}, \citenamefont {Pitsillides}, \citenamefont
  {Akyildiz}, \citenamefont {Kantartzis}, \citenamefont {Economou},
  \citenamefont {Soukoulis}, \citenamefont {Kafesaki},\ and\ \citenamefont
  {Tretyakov}}]{Tsilipakos2020progress}%
  \BibitemOpen
  \bibfield  {author} {\bibinfo {author} {\bibfnamefont {O.}~\bibnamefont
  {Tsilipakos}}, \bibinfo {author} {\bibfnamefont {A.~C.}\ \bibnamefont
  {Tasolamprou}}, \bibinfo {author} {\bibfnamefont {A.}~\bibnamefont
  {Pitilakis}}, \bibinfo {author} {\bibfnamefont {F.}~\bibnamefont {Liu}},
  \bibinfo {author} {\bibfnamefont {X.}~\bibnamefont {Wang}}, \bibinfo {author}
  {\bibfnamefont {M.~S.}\ \bibnamefont {Mirmoosa}}, \bibinfo {author}
  {\bibfnamefont {D.~C.}\ \bibnamefont {Tzarouchis}}, \bibinfo {author}
  {\bibfnamefont {S.}~\bibnamefont {Abadal}}, \bibinfo {author} {\bibfnamefont
  {H.}~\bibnamefont {Taghvaee}}, \bibinfo {author} {\bibfnamefont
  {C.}~\bibnamefont {Liaskos}}, \bibinfo {author} {\bibfnamefont
  {A.}~\bibnamefont {Tsioliaridou}}, \bibinfo {author} {\bibfnamefont
  {J.}~\bibnamefont {Georgiou}}, \bibinfo {author} {\bibfnamefont
  {A.}~\bibnamefont {Cabellos-Aparicio}}, \bibinfo {author} {\bibfnamefont
  {E.}~\bibnamefont {Alarc{\'{o}}n}}, \bibinfo {author} {\bibfnamefont
  {S.}~\bibnamefont {Ioannidis}}, \bibinfo {author} {\bibfnamefont
  {A.}~\bibnamefont {Pitsillides}}, \bibinfo {author} {\bibfnamefont {I.~F.}\
  \bibnamefont {Akyildiz}}, \bibinfo {author} {\bibfnamefont {N.~V.}\
  \bibnamefont {Kantartzis}}, \bibinfo {author} {\bibfnamefont {E.~N.}\
  \bibnamefont {Economou}}, \bibinfo {author} {\bibfnamefont {C.~M.}\
  \bibnamefont {Soukoulis}}, \bibinfo {author} {\bibfnamefont {M.}~\bibnamefont
  {Kafesaki}},\ and\ \bibinfo {author} {\bibfnamefont {S.}~\bibnamefont
  {Tretyakov}},\ }\bibfield  {title} {\bibinfo {title} {Toward intelligent
  metasurfaces: The progress from globally tunable metasurfaces to
  software-defined metasurfaces with an embedded network of controllers},\
  }\href {https://doi.org/10.1002/adom.202000783} {\bibfield  {journal}
  {\bibinfo  {journal} {Advanced Optical Materials}\ }\textbf {\bibinfo
  {volume} {8}},\ \bibinfo {pages} {2000783} (\bibinfo {year}
  {2020})}\BibitemShut {NoStop}%
\bibitem [{\citenamefont {Liu}\ \emph {et~al.}(2019)\citenamefont {Liu},
  \citenamefont {Tsilipakos}, \citenamefont {Pitilakis}, \citenamefont
  {Tasolamprou}, \citenamefont {Mirmoosa}, \citenamefont {Kantartzis},
  \citenamefont {Kwon}, \citenamefont {Georgiou}, \citenamefont {Kossifos},
  \citenamefont {Antoniades}, \citenamefont {Kafesaki}, \citenamefont
  {Soukoulis},\ and\ \citenamefont {Tretyakov}}]{Liu:2019}%
  \BibitemOpen
  \bibfield  {author} {\bibinfo {author} {\bibfnamefont {F.}~\bibnamefont
  {Liu}}, \bibinfo {author} {\bibfnamefont {O.}~\bibnamefont {Tsilipakos}},
  \bibinfo {author} {\bibfnamefont {A.}~\bibnamefont {Pitilakis}}, \bibinfo
  {author} {\bibfnamefont {A.~C.}\ \bibnamefont {Tasolamprou}}, \bibinfo
  {author} {\bibfnamefont {M.~S.}\ \bibnamefont {Mirmoosa}}, \bibinfo {author}
  {\bibfnamefont {N.~V.}\ \bibnamefont {Kantartzis}}, \bibinfo {author}
  {\bibfnamefont {D.-H.}\ \bibnamefont {Kwon}}, \bibinfo {author}
  {\bibfnamefont {J.}~\bibnamefont {Georgiou}}, \bibinfo {author}
  {\bibfnamefont {K.}~\bibnamefont {Kossifos}}, \bibinfo {author}
  {\bibfnamefont {M.~A.}\ \bibnamefont {Antoniades}}, \bibinfo {author}
  {\bibfnamefont {M.}~\bibnamefont {Kafesaki}}, \bibinfo {author}
  {\bibfnamefont {C.~M.}\ \bibnamefont {Soukoulis}},\ and\ \bibinfo {author}
  {\bibfnamefont {S.~A.}\ \bibnamefont {Tretyakov}},\ }\bibfield  {title}
  {\bibinfo {title} {Intelligent metasurfaces with continuously tunable local
  surface impedance for multiple reconfigurable functions},\ }\href@noop {}
  {\bibfield  {journal} {\bibinfo  {journal} {Phys. Rev. Appl.}\ }\textbf
  {\bibinfo {volume} {11}},\ \bibinfo {pages} {044024} (\bibinfo {year}
  {2019})}\BibitemShut {NoStop}%
\bibitem [{\citenamefont {Pitilakis}\ \emph {et~al.}(2018)\citenamefont
  {Pitilakis}, \citenamefont {Tasolamprou}, \citenamefont {Liaskos},
  \citenamefont {Liu}, \citenamefont {Tsilipakos}, \citenamefont {Wang},
  \citenamefont {Mirmoosa}, \citenamefont {Kossifos}, \citenamefont {Georgiou},
  \citenamefont {Pitsilides}, \citenamefont {Kantartzis}, \citenamefont
  {Ioannidis}, \citenamefont {Economou}, \citenamefont {Kafesaki},
  \citenamefont {Tretyakov},\ and\ \citenamefont {Soukoulis}}]{Pitilakis:2018}%
  \BibitemOpen
  \bibfield  {author} {\bibinfo {author} {\bibfnamefont {A.}~\bibnamefont
  {Pitilakis}}, \bibinfo {author} {\bibfnamefont {A.~C.}\ \bibnamefont
  {Tasolamprou}}, \bibinfo {author} {\bibfnamefont {C.}~\bibnamefont
  {Liaskos}}, \bibinfo {author} {\bibfnamefont {F.}~\bibnamefont {Liu}},
  \bibinfo {author} {\bibfnamefont {O.}~\bibnamefont {Tsilipakos}}, \bibinfo
  {author} {\bibfnamefont {X.}~\bibnamefont {Wang}}, \bibinfo {author}
  {\bibfnamefont {M.~S.}\ \bibnamefont {Mirmoosa}}, \bibinfo {author}
  {\bibfnamefont {K.}~\bibnamefont {Kossifos}}, \bibinfo {author}
  {\bibfnamefont {J.}~\bibnamefont {Georgiou}}, \bibinfo {author}
  {\bibfnamefont {A.}~\bibnamefont {Pitsilides}}, \bibinfo {author}
  {\bibfnamefont {N.}~\bibnamefont {Kantartzis}}, \bibinfo {author}
  {\bibfnamefont {S.}~\bibnamefont {Ioannidis}}, \bibinfo {author}
  {\bibfnamefont {E.~N.}\ \bibnamefont {Economou}}, \bibinfo {author}
  {\bibfnamefont {M.}~\bibnamefont {Kafesaki}}, \bibinfo {author}
  {\bibfnamefont {S.~A.}\ \bibnamefont {Tretyakov}},\ and\ \bibinfo {author}
  {\bibfnamefont {C.~M.}\ \bibnamefont {Soukoulis}},\ }\bibfield  {title}
  {\bibinfo {title} {Software-defined metasurface paradigm: Concept,
  challenges, prospects},\ }in\ \href
  {https://doi.org/10.1109/metamaterials.2018.8534096} {\emph {\bibinfo
  {booktitle} {Metamaterials 2018}}}\ (\bibinfo  {publisher} {IEEE},\ \bibinfo
  {year} {2018})\BibitemShut {NoStop}%
\bibitem [{\citenamefont {Assimonis}\ and\ \citenamefont
  {Fusco}(2019)}]{Assimonis2019}%
  \BibitemOpen
  \bibfield  {author} {\bibinfo {author} {\bibfnamefont {S.~D.}\ \bibnamefont
  {Assimonis}}\ and\ \bibinfo {author} {\bibfnamefont {V.}~\bibnamefont
  {Fusco}},\ }\bibfield  {title} {\bibinfo {title} {Polarization insensitive,
  wide-angle, ultra-wideband, flexible, resistively loaded, electromagnetic
  metamaterial absorber using conventional inkjet-printing technology},\
  }\bibfield  {journal} {\bibinfo  {journal} {Scientific Reports}\ }\textbf
  {\bibinfo {volume} {9}},\ \href {https://doi.org/10.1038/s41598-019-48761-6}
  {10.1038/s41598-019-48761-6} (\bibinfo {year} {2019})\BibitemShut {NoStop}%
\bibitem [{\citenamefont {Ramaccia}\ \emph {et~al.}(2020)\citenamefont
  {Ramaccia}, \citenamefont {Sounas}, \citenamefont {Marini}, \citenamefont
  {Toscano},\ and\ \citenamefont {Bilotti}}]{Ramaccia2020}%
  \BibitemOpen
  \bibfield  {author} {\bibinfo {author} {\bibfnamefont {D.}~\bibnamefont
  {Ramaccia}}, \bibinfo {author} {\bibfnamefont {D.~L.}\ \bibnamefont
  {Sounas}}, \bibinfo {author} {\bibfnamefont {A.~V.}\ \bibnamefont {Marini}},
  \bibinfo {author} {\bibfnamefont {A.}~\bibnamefont {Toscano}},\ and\ \bibinfo
  {author} {\bibfnamefont {F.}~\bibnamefont {Bilotti}},\ }\bibfield  {title}
  {\bibinfo {title} {Electromagnetic isolation induced by time-varying
  metasurfaces: Nonreciprocal bragg grating},\ }\href
  {https://doi.org/10.1109/lawp.2020.2996275} {\bibfield  {journal} {\bibinfo
  {journal} {{IEEE} {AWPL}}\ }\textbf {\bibinfo {volume} {19}},\ \bibinfo
  {pages} {1886} (\bibinfo {year} {2020})}\BibitemShut {NoStop}%
\bibitem [{\citenamefont {Taravati}\ and\ \citenamefont
  {Eleftheriades}(2020)}]{Taravati2020}%
  \BibitemOpen
  \bibfield  {author} {\bibinfo {author} {\bibfnamefont {S.}~\bibnamefont
  {Taravati}}\ and\ \bibinfo {author} {\bibfnamefont {G.~V.}\ \bibnamefont
  {Eleftheriades}},\ }\bibfield  {title} {\bibinfo {title} {Full-duplex
  nonreciprocal beam steering by time-modulated phase-gradient metasurfaces},\
  }\href {https://doi.org/10.1103/physrevapplied.14.014027} {\bibfield
  {journal} {\bibinfo  {journal} {Physical Review Applied}\ }\textbf {\bibinfo
  {volume} {14}},\ \bibinfo {pages} {014027} (\bibinfo {year}
  {2020})}\BibitemShut {NoStop}%
\bibitem [{\citenamefont {Hwang}\ \emph {et~al.}(2021)\citenamefont {Hwang},
  \citenamefont {Yun}, \citenamefont {Park}, \citenamefont {Hong},\ and\
  \citenamefont {Lee}}]{Hwang2021}%
  \BibitemOpen
  \bibfield  {author} {\bibinfo {author} {\bibfnamefont {I.-J.}\ \bibnamefont
  {Hwang}}, \bibinfo {author} {\bibfnamefont {D.-J.}\ \bibnamefont {Yun}},
  \bibinfo {author} {\bibfnamefont {J.-I.}\ \bibnamefont {Park}}, \bibinfo
  {author} {\bibfnamefont {Y.-P.}\ \bibnamefont {Hong}},\ and\ \bibinfo
  {author} {\bibfnamefont {I.-H.}\ \bibnamefont {Lee}},\ }\bibfield  {title}
  {\bibinfo {title} {Design of dual-band single-layer metasurfaces for
  millimeter-wave {5G} communication systems},\ }\href
  {https://doi.org/10.1063/5.0062064} {\bibfield  {journal} {\bibinfo
  {journal} {Applied Physics Letters}\ }\textbf {\bibinfo {volume} {119}},\
  \bibinfo {pages} {174101} (\bibinfo {year} {2021})}\BibitemShut {NoStop}%
\bibitem [{\citenamefont {de~Lustrac}\ \emph {et~al.}(2021)\citenamefont
  {de~Lustrac}, \citenamefont {Ratni}, \citenamefont {Piau}, \citenamefont
  {Duval},\ and\ \citenamefont {Burokur}}]{deLustrac2021}%
  \BibitemOpen
  \bibfield  {author} {\bibinfo {author} {\bibfnamefont {A.}~\bibnamefont
  {de~Lustrac}}, \bibinfo {author} {\bibfnamefont {B.}~\bibnamefont {Ratni}},
  \bibinfo {author} {\bibfnamefont {G.-P.}\ \bibnamefont {Piau}}, \bibinfo
  {author} {\bibfnamefont {Y.}~\bibnamefont {Duval}},\ and\ \bibinfo {author}
  {\bibfnamefont {S.~N.}\ \bibnamefont {Burokur}},\ }\bibfield  {title}
  {\bibinfo {title} {Tri-state metasurface-based electromagnetic screen with
  switchable reflection, transmission, and absorption functionalities},\ }\href
  {https://doi.org/10.1021/acsaelm.0c01038} {\bibfield  {journal} {\bibinfo
  {journal} {{ACS} Applied Electronic Materials}\ }\textbf {\bibinfo {volume}
  {3}},\ \bibinfo {pages} {1184} (\bibinfo {year} {2021})}\BibitemShut
  {NoStop}%
\bibitem [{\citenamefont {Smith}\ \emph {et~al.}(2017)\citenamefont {Smith},
  \citenamefont {Yurduseven}, \citenamefont {Mancera}, \citenamefont {Bowen},\
  and\ \citenamefont {Kundtz}}]{Smith2017}%
  \BibitemOpen
  \bibfield  {author} {\bibinfo {author} {\bibfnamefont {D.~R.}\ \bibnamefont
  {Smith}}, \bibinfo {author} {\bibfnamefont {O.}~\bibnamefont {Yurduseven}},
  \bibinfo {author} {\bibfnamefont {L.~P.}\ \bibnamefont {Mancera}}, \bibinfo
  {author} {\bibfnamefont {P.}~\bibnamefont {Bowen}},\ and\ \bibinfo {author}
  {\bibfnamefont {N.~B.}\ \bibnamefont {Kundtz}},\ }\bibfield  {title}
  {\bibinfo {title} {Analysis of a waveguide-fed metasurface antenna},\ }\href
  {https://doi.org/10.1103/physrevapplied.8.054048} {\bibfield  {journal}
  {\bibinfo  {journal} {Physical Review Applied}\ }\textbf {\bibinfo {volume}
  {8}},\ \bibinfo {pages} {054048} (\bibinfo {year} {2017})}\BibitemShut
  {NoStop}%
\bibitem [{\citenamefont {Yurduseven}\ \emph {et~al.}(2020)\citenamefont
  {Yurduseven}, \citenamefont {Assimonis},\ and\ \citenamefont
  {Matthaiou}}]{Yurduseven2020}%
  \BibitemOpen
  \bibfield  {author} {\bibinfo {author} {\bibfnamefont {O.}~\bibnamefont
  {Yurduseven}}, \bibinfo {author} {\bibfnamefont {S.~D.}\ \bibnamefont
  {Assimonis}},\ and\ \bibinfo {author} {\bibfnamefont {M.}~\bibnamefont
  {Matthaiou}},\ }\bibfield  {title} {\bibinfo {title} {Intelligent reflecting
  surfaces with spatial modulation: An electromagnetic perspective},\ }\href
  {https://doi.org/10.1109/ojcoms.2020.3017237} {\bibfield  {journal} {\bibinfo
   {journal} {{IEEE} Open J. Comm. Soc.}\ }\textbf {\bibinfo {volume} {1}},\
  \bibinfo {pages} {1256} (\bibinfo {year} {2020})}\BibitemShut {NoStop}%
\bibitem [{\citenamefont {Cai}\ \emph {et~al.}(2017)\citenamefont {Cai},
  \citenamefont {Wang}, \citenamefont {Tang}, \citenamefont {Xu}, \citenamefont
  {Duan}, \citenamefont {Guo}, \citenamefont {Guan}, \citenamefont {Sun},
  \citenamefont {He},\ and\ \citenamefont {Zhou}}]{Cai2017}%
  \BibitemOpen
  \bibfield  {author} {\bibinfo {author} {\bibfnamefont {T.}~\bibnamefont
  {Cai}}, \bibinfo {author} {\bibfnamefont {G.}~\bibnamefont {Wang}}, \bibinfo
  {author} {\bibfnamefont {S.}~\bibnamefont {Tang}}, \bibinfo {author}
  {\bibfnamefont {H.}~\bibnamefont {Xu}}, \bibinfo {author} {\bibfnamefont
  {J.}~\bibnamefont {Duan}}, \bibinfo {author} {\bibfnamefont {H.}~\bibnamefont
  {Guo}}, \bibinfo {author} {\bibfnamefont {F.}~\bibnamefont {Guan}}, \bibinfo
  {author} {\bibfnamefont {S.}~\bibnamefont {Sun}}, \bibinfo {author}
  {\bibfnamefont {Q.}~\bibnamefont {He}},\ and\ \bibinfo {author}
  {\bibfnamefont {L.}~\bibnamefont {Zhou}},\ }\bibfield  {title} {\bibinfo
  {title} {High-efficiency and full-space manipulation of electromagnetic wave
  fronts with metasurfaces},\ }\href
  {https://doi.org/10.1103/physrevapplied.8.034033} {\bibfield  {journal}
  {\bibinfo  {journal} {Physical Review Applied}\ }\textbf {\bibinfo {volume}
  {8}},\ \bibinfo {pages} {034033} (\bibinfo {year} {2017})}\BibitemShut
  {NoStop}%
\bibitem [{\citenamefont {Feng}\ \emph {et~al.}(2021)\citenamefont {Feng},
  \citenamefont {Ratni}, \citenamefont {Yi}, \citenamefont {Zhang},
  \citenamefont {de~Lustrac},\ and\ \citenamefont {Burokur}}]{Feng2021}%
  \BibitemOpen
  \bibfield  {author} {\bibinfo {author} {\bibfnamefont {R.}~\bibnamefont
  {Feng}}, \bibinfo {author} {\bibfnamefont {B.}~\bibnamefont {Ratni}},
  \bibinfo {author} {\bibfnamefont {J.}~\bibnamefont {Yi}}, \bibinfo {author}
  {\bibfnamefont {H.}~\bibnamefont {Zhang}}, \bibinfo {author} {\bibfnamefont
  {A.}~\bibnamefont {de~Lustrac}},\ and\ \bibinfo {author} {\bibfnamefont
  {S.~N.}\ \bibnamefont {Burokur}},\ }\bibfield  {title} {\bibinfo {title}
  {Design and experimental characterization of a two-dimensional reconfigurable
  metasurface},\ }in\ \href {https://doi.org/10.23919/eucap51087.2021.9411004}
  {\emph {\bibinfo {booktitle} {2021 {EuCAP}}}}\ (\bibinfo  {publisher}
  {{IEEE}},\ \bibinfo {year} {2021})\BibitemShut {NoStop}%
\bibitem [{\citenamefont {Pitilakis}\ \emph {et~al.}(2021)\citenamefont
  {Pitilakis}, \citenamefont {Tsilipakos}, \citenamefont {Liu}, \citenamefont
  {Kossifos}, \citenamefont {Tasolamprou}, \citenamefont {Kwon}, \citenamefont
  {Mirmoosa}, \citenamefont {Manessis}, \citenamefont {Kantartzis},
  \citenamefont {Liaskos}, \citenamefont {Antoniades}, \citenamefont
  {Georgiou}, \citenamefont {Soukoulis}, \citenamefont {Kafesaki},\ and\
  \citenamefont {Tretyakov}}]{Pitilakis2021}%
  \BibitemOpen
  \bibfield  {author} {\bibinfo {author} {\bibfnamefont {A.}~\bibnamefont
  {Pitilakis}}, \bibinfo {author} {\bibfnamefont {O.}~\bibnamefont
  {Tsilipakos}}, \bibinfo {author} {\bibfnamefont {F.}~\bibnamefont {Liu}},
  \bibinfo {author} {\bibfnamefont {K.~M.}\ \bibnamefont {Kossifos}}, \bibinfo
  {author} {\bibfnamefont {A.~C.}\ \bibnamefont {Tasolamprou}}, \bibinfo
  {author} {\bibfnamefont {D.-H.}\ \bibnamefont {Kwon}}, \bibinfo {author}
  {\bibfnamefont {M.~S.}\ \bibnamefont {Mirmoosa}}, \bibinfo {author}
  {\bibfnamefont {D.}~\bibnamefont {Manessis}}, \bibinfo {author}
  {\bibfnamefont {N.~V.}\ \bibnamefont {Kantartzis}}, \bibinfo {author}
  {\bibfnamefont {C.}~\bibnamefont {Liaskos}}, \bibinfo {author} {\bibfnamefont
  {M.~A.}\ \bibnamefont {Antoniades}}, \bibinfo {author} {\bibfnamefont
  {J.}~\bibnamefont {Georgiou}}, \bibinfo {author} {\bibfnamefont {C.~M.}\
  \bibnamefont {Soukoulis}}, \bibinfo {author} {\bibfnamefont {M.}~\bibnamefont
  {Kafesaki}},\ and\ \bibinfo {author} {\bibfnamefont {S.~A.}\ \bibnamefont
  {Tretyakov}},\ }\bibfield  {title} {\bibinfo {title} {A multi-functional
  reconfigurable metasurface: Electromagnetic design accounting for fabrication
  aspects},\ }\href {https://doi.org/10.1109/tap.2020.3016479} {\bibfield
  {journal} {\bibinfo  {journal} {{IEEE} Trans. Antennas Propag.}\ }\textbf
  {\bibinfo {volume} {69}},\ \bibinfo {pages} {1440} (\bibinfo {year}
  {2021})}\BibitemShut {NoStop}%
\bibitem [{\citenamefont {Kossifos}\ \emph {et~al.}(2020)\citenamefont
  {Kossifos}, \citenamefont {Petrou}, \citenamefont {Varnava}, \citenamefont
  {Pitilakis}, \citenamefont {Tsilipakos}, \citenamefont {Liu}, \citenamefont
  {Karousios}, \citenamefont {Tasolamprou}, \citenamefont {Seckel},
  \citenamefont {Manessis}, \citenamefont {Kantartzis}, \citenamefont {Kwon},
  \citenamefont {Antoniades},\ and\ \citenamefont {Georgiou}}]{Kossifos2020}%
  \BibitemOpen
  \bibfield  {author} {\bibinfo {author} {\bibfnamefont {K.~M.}\ \bibnamefont
  {Kossifos}}, \bibinfo {author} {\bibfnamefont {L.}~\bibnamefont {Petrou}},
  \bibinfo {author} {\bibfnamefont {G.}~\bibnamefont {Varnava}}, \bibinfo
  {author} {\bibfnamefont {A.}~\bibnamefont {Pitilakis}}, \bibinfo {author}
  {\bibfnamefont {O.}~\bibnamefont {Tsilipakos}}, \bibinfo {author}
  {\bibfnamefont {F.}~\bibnamefont {Liu}}, \bibinfo {author} {\bibfnamefont
  {P.}~\bibnamefont {Karousios}}, \bibinfo {author} {\bibfnamefont {A.~C.}\
  \bibnamefont {Tasolamprou}}, \bibinfo {author} {\bibfnamefont
  {M.}~\bibnamefont {Seckel}}, \bibinfo {author} {\bibfnamefont
  {D.}~\bibnamefont {Manessis}}, \bibinfo {author} {\bibfnamefont {N.~V.}\
  \bibnamefont {Kantartzis}}, \bibinfo {author} {\bibfnamefont {D.-H.}\
  \bibnamefont {Kwon}}, \bibinfo {author} {\bibfnamefont {M.~A.}\ \bibnamefont
  {Antoniades}},\ and\ \bibinfo {author} {\bibfnamefont {J.}~\bibnamefont
  {Georgiou}},\ }\bibfield  {title} {\bibinfo {title} {Toward the realization
  of a programmable metasurface absorber enabled by custom integrated circuit
  technology},\ }\href {https://doi.org/10.1109/access.2020.2994469} {\bibfield
   {journal} {\bibinfo  {journal} {{IEEE} Access}\ }\textbf {\bibinfo {volume}
  {8}},\ \bibinfo {pages} {92986} (\bibinfo {year} {2020})}\BibitemShut
  {NoStop}%
\bibitem [{\citenamefont {Koziel}\ and\ \citenamefont
  {Pietrenko-Dabrowska}(2021)}]{Koziel2021}%
  \BibitemOpen
  \bibfield  {author} {\bibinfo {author} {\bibfnamefont {S.}~\bibnamefont
  {Koziel}}\ and\ \bibinfo {author} {\bibfnamefont {A.}~\bibnamefont
  {Pietrenko-Dabrowska}},\ }\bibfield  {title} {\bibinfo {title} {On geometry
  parameterization for simulation-driven design closure of antenna
  structures},\ }\bibfield  {journal} {\bibinfo  {journal} {Scientific
  Reports}\ }\textbf {\bibinfo {volume} {11}},\ \href
  {https://doi.org/10.1038/s41598-021-03728-4} {10.1038/s41598-021-03728-4}
  (\bibinfo {year} {2021})\BibitemShut {NoStop}%
\bibitem [{\citenamefont {Manessis}\ \emph {et~al.}(2020)\citenamefont
  {Manessis}, \citenamefont {Seckel}, \citenamefont {Fu}, \citenamefont
  {Tsilipakos}, \citenamefont {Pitilakis}, \citenamefont {Tasolamprou},
  \citenamefont {Kossifos}, \citenamefont {Varnava}, \citenamefont {Liaskos},
  \citenamefont {Kafesaki}, \citenamefont {Soukoulis}, \citenamefont
  {Tretyakov}, \citenamefont {Georgiou}, \citenamefont {Ostmann}, \citenamefont
  {Aschenbrenner}, \citenamefont {Schneider-Ramelow},\ and\ \citenamefont
  {Lang}}]{Manessis2020}%
  \BibitemOpen
  \bibfield  {author} {\bibinfo {author} {\bibfnamefont {D.}~\bibnamefont
  {Manessis}}, \bibinfo {author} {\bibfnamefont {M.}~\bibnamefont {Seckel}},
  \bibinfo {author} {\bibfnamefont {L.}~\bibnamefont {Fu}}, \bibinfo {author}
  {\bibfnamefont {O.}~\bibnamefont {Tsilipakos}}, \bibinfo {author}
  {\bibfnamefont {A.}~\bibnamefont {Pitilakis}}, \bibinfo {author}
  {\bibfnamefont {A.}~\bibnamefont {Tasolamprou}}, \bibinfo {author}
  {\bibfnamefont {K.}~\bibnamefont {Kossifos}}, \bibinfo {author}
  {\bibfnamefont {G.}~\bibnamefont {Varnava}}, \bibinfo {author} {\bibfnamefont
  {C.}~\bibnamefont {Liaskos}}, \bibinfo {author} {\bibfnamefont
  {M.}~\bibnamefont {Kafesaki}}, \bibinfo {author} {\bibfnamefont {C.~M.}\
  \bibnamefont {Soukoulis}}, \bibinfo {author} {\bibfnamefont {S.}~\bibnamefont
  {Tretyakov}}, \bibinfo {author} {\bibfnamefont {J.}~\bibnamefont {Georgiou}},
  \bibinfo {author} {\bibfnamefont {A.}~\bibnamefont {Ostmann}}, \bibinfo
  {author} {\bibfnamefont {R.}~\bibnamefont {Aschenbrenner}}, \bibinfo {author}
  {\bibfnamefont {M.}~\bibnamefont {Schneider-Ramelow}},\ and\ \bibinfo
  {author} {\bibfnamefont {K.-D.}\ \bibnamefont {Lang}},\ }\bibfield  {title}
  {\bibinfo {title} {Manufacturing of high frequency substrates as software
  programmable metasurfaces on {PCBs} with integrated controller nodes},\ }in\
  \href {https://doi.org/10.1109/estc48849.2020.9229660} {\emph {\bibinfo
  {booktitle} {2020 {IEEE} {ESTC}}}}\ (\bibinfo  {publisher} {{IEEE}},\
  \bibinfo {year} {2020})\BibitemShut {NoStop}%
\bibitem [{\citenamefont {Manessis}\ \emph {et~al.}(2021)\citenamefont
  {Manessis}, \citenamefont {Kosmider}, \citenamefont {Boettcher},
  \citenamefont {Seckel}, \citenamefont {Murugesan}, \citenamefont {Maas},
  \citenamefont {Ndip}, \citenamefont {Ostmann}, \citenamefont {Aschenbrenner},
  \citenamefont {Schneider-Ramelow},\ and\ \citenamefont
  {Lang}}]{Manessis2021}%
  \BibitemOpen
  \bibfield  {author} {\bibinfo {author} {\bibfnamefont {D.}~\bibnamefont
  {Manessis}}, \bibinfo {author} {\bibfnamefont {S.}~\bibnamefont {Kosmider}},
  \bibinfo {author} {\bibfnamefont {L.}~\bibnamefont {Boettcher}}, \bibinfo
  {author} {\bibfnamefont {M.}~\bibnamefont {Seckel}}, \bibinfo {author}
  {\bibfnamefont {K.}~\bibnamefont {Murugesan}}, \bibinfo {author}
  {\bibfnamefont {U.}~\bibnamefont {Maas}}, \bibinfo {author} {\bibfnamefont
  {I.}~\bibnamefont {Ndip}}, \bibinfo {author} {\bibfnamefont {A.}~\bibnamefont
  {Ostmann}}, \bibinfo {author} {\bibfnamefont {R.}~\bibnamefont
  {Aschenbrenner}}, \bibinfo {author} {\bibfnamefont {M.}~\bibnamefont
  {Schneider-Ramelow}},\ and\ \bibinfo {author} {\bibfnamefont {K.-D.}\
  \bibnamefont {Lang}},\ }\bibfield  {title} {\bibinfo {title} {Development of
  innovative substrate and embedding technologies for high frequency
  applications},\ }in\ \href {https://doi.org/10.23919/empc53418.2021.9584975}
  {\emph {\bibinfo {booktitle} {2021 {EMPC}}}}\ (\bibinfo  {publisher}
  {{IEEE}},\ \bibinfo {year} {2021})\BibitemShut {NoStop}%
\bibitem [{Note1()}]{Note1}%
  \BibitemOpen
  \bibinfo {note} {See the Supplemental Material for (i)~conceptual strategy
  for multi-functionality, (ii)~implementation details for full-wave simulation
  at the unit-cell level, (iii)~additional photos of fabricated samples and
  measurement setup, (iv)~simplified model and underlying resonances of
  polarization conversion functionality, (v)~cross-polarization response for
  oblique incidence, (vi)~unit-cell states for beam splitting functionality,
  (vii)~details of Huygens-Fresnel methodology for obtaining the farfield
  radiation pattern.}\BibitemShut {Stop}%
\bibitem [{\citenamefont {Zhang}\ \emph {et~al.}(2020)\citenamefont {Zhang},
  \citenamefont {Li}, \citenamefont {Liu}, \citenamefont {Qiu}, \citenamefont
  {Sun}, \citenamefont {He},\ and\ \citenamefont {Zhou}}]{Zhang2020}%
  \BibitemOpen
  \bibfield  {author} {\bibinfo {author} {\bibfnamefont {X.}~\bibnamefont
  {Zhang}}, \bibinfo {author} {\bibfnamefont {Q.}~\bibnamefont {Li}}, \bibinfo
  {author} {\bibfnamefont {F.}~\bibnamefont {Liu}}, \bibinfo {author}
  {\bibfnamefont {M.}~\bibnamefont {Qiu}}, \bibinfo {author} {\bibfnamefont
  {S.}~\bibnamefont {Sun}}, \bibinfo {author} {\bibfnamefont {Q.}~\bibnamefont
  {He}},\ and\ \bibinfo {author} {\bibfnamefont {L.}~\bibnamefont {Zhou}},\
  }\bibfield  {title} {\bibinfo {title} {Controlling angular dispersions in
  optical metasurfaces},\ }\bibfield  {journal} {\bibinfo  {journal} {Light:
  Science {\&} Applications}\ }\textbf {\bibinfo {volume} {9}},\ \href
  {https://doi.org/10.1038/s41377-020-0313-0} {10.1038/s41377-020-0313-0}
  (\bibinfo {year} {2020})\BibitemShut {NoStop}%
\bibitem [{\citenamefont {Christopoulos}\ \emph {et~al.}(2019)\citenamefont
  {Christopoulos}, \citenamefont {Tsilipakos}, \citenamefont {Sinatkas},\ and\
  \citenamefont {Kriezis}}]{Christopoulos2019}%
  \BibitemOpen
  \bibfield  {author} {\bibinfo {author} {\bibfnamefont {T.}~\bibnamefont
  {Christopoulos}}, \bibinfo {author} {\bibfnamefont {O.}~\bibnamefont
  {Tsilipakos}}, \bibinfo {author} {\bibfnamefont {G.}~\bibnamefont
  {Sinatkas}},\ and\ \bibinfo {author} {\bibfnamefont {E.~E.}\ \bibnamefont
  {Kriezis}},\ }\bibfield  {title} {\bibinfo {title} {On the calculation of the
  quality factor in contemporary photonic resonant structures},\ }\href
  {https://doi.org/10.1364/OE.27.014505} {\bibfield  {journal} {\bibinfo
  {journal} {Opt. Express}\ }\textbf {\bibinfo {volume} {27}},\ \bibinfo
  {pages} {14505} (\bibinfo {year} {2019})}\BibitemShut {NoStop}%
\bibitem [{\citenamefont {Grady}\ \emph {et~al.}(2013)\citenamefont {Grady},
  \citenamefont {Heyes}, \citenamefont {Chowdhury}, \citenamefont {Zeng},
  \citenamefont {Reiten}, \citenamefont {Azad}, \citenamefont {Taylor},
  \citenamefont {Dalvit},\ and\ \citenamefont {Chen}}]{Grady:2013}%
  \BibitemOpen
  \bibfield  {author} {\bibinfo {author} {\bibfnamefont {N.~K.}\ \bibnamefont
  {Grady}}, \bibinfo {author} {\bibfnamefont {J.~E.}\ \bibnamefont {Heyes}},
  \bibinfo {author} {\bibfnamefont {D.~R.}\ \bibnamefont {Chowdhury}}, \bibinfo
  {author} {\bibfnamefont {Y.}~\bibnamefont {Zeng}}, \bibinfo {author}
  {\bibfnamefont {M.~T.}\ \bibnamefont {Reiten}}, \bibinfo {author}
  {\bibfnamefont {A.~K.}\ \bibnamefont {Azad}}, \bibinfo {author}
  {\bibfnamefont {A.~J.}\ \bibnamefont {Taylor}}, \bibinfo {author}
  {\bibfnamefont {D.~A.~R.}\ \bibnamefont {Dalvit}},\ and\ \bibinfo {author}
  {\bibfnamefont {H.~T.}\ \bibnamefont {Chen}},\ }\bibfield  {title} {\bibinfo
  {title} {Terahertz metamaterials for linear polarization conversion and
  anomalous refraction},\ }\href@noop {} {\bibfield  {journal} {\bibinfo
  {journal} {Science}\ }\textbf {\bibinfo {volume} {304}},\ \bibinfo {pages}
  {1304} (\bibinfo {year} {2013})}\BibitemShut {NoStop}%
\bibitem [{\citenamefont {L{\'{e}}vesque}\ \emph {et~al.}(2014)\citenamefont
  {L{\'{e}}vesque}, \citenamefont {Makhsiyan}, \citenamefont {Bouchon},
  \citenamefont {Pardo}, \citenamefont {Jaeck}, \citenamefont {Bardou},
  \citenamefont {Dupuis}, \citenamefont {Haïdar},\ and\ \citenamefont
  {Pelouard}}]{Levesque2014}%
  \BibitemOpen
  \bibfield  {author} {\bibinfo {author} {\bibfnamefont {Q.}~\bibnamefont
  {L{\'{e}}vesque}}, \bibinfo {author} {\bibfnamefont {M.}~\bibnamefont
  {Makhsiyan}}, \bibinfo {author} {\bibfnamefont {P.}~\bibnamefont {Bouchon}},
  \bibinfo {author} {\bibfnamefont {F.}~\bibnamefont {Pardo}}, \bibinfo
  {author} {\bibfnamefont {J.}~\bibnamefont {Jaeck}}, \bibinfo {author}
  {\bibfnamefont {N.}~\bibnamefont {Bardou}}, \bibinfo {author} {\bibfnamefont
  {C.}~\bibnamefont {Dupuis}}, \bibinfo {author} {\bibfnamefont
  {R.}~\bibnamefont {Haïdar}},\ and\ \bibinfo {author} {\bibfnamefont {J.-L.}\
  \bibnamefont {Pelouard}},\ }\bibfield  {title} {\bibinfo {title} {Plasmonic
  planar antenna for wideband and efficient linear polarization conversion},\
  }\href {https://doi.org/10.1063/1.4869127} {\bibfield  {journal} {\bibinfo
  {journal} {Applied Physics Letters}\ }\textbf {\bibinfo {volume} {104}},\
  \bibinfo {pages} {111105} (\bibinfo {year} {2014})}\BibitemShut {NoStop}%
\bibitem [{\citenamefont {Tsilipakos}\ \emph {et~al.}(2021)\citenamefont
  {Tsilipakos}, \citenamefont {Zhang}, \citenamefont {Kafesaki}, \citenamefont
  {Soukoulis},\ and\ \citenamefont {Koschny}}]{Tsilipakos2021}%
  \BibitemOpen
  \bibfield  {author} {\bibinfo {author} {\bibfnamefont {O.}~\bibnamefont
  {Tsilipakos}}, \bibinfo {author} {\bibfnamefont {L.}~\bibnamefont {Zhang}},
  \bibinfo {author} {\bibfnamefont {M.}~\bibnamefont {Kafesaki}}, \bibinfo
  {author} {\bibfnamefont {C.~M.}\ \bibnamefont {Soukoulis}},\ and\ \bibinfo
  {author} {\bibfnamefont {T.}~\bibnamefont {Koschny}},\ }\bibfield  {title}
  {\bibinfo {title} {Experimental implementation of achromatic multiresonant
  metasurface for broadband pulse delay},\ }\href
  {https://doi.org/10.1021/acsphotonics.1c00025} {\bibfield  {journal}
  {\bibinfo  {journal} {{ACS} Photonics}\ }\textbf {\bibinfo {volume} {8}},\
  \bibinfo {pages} {1649} (\bibinfo {year} {2021})}\BibitemShut {NoStop}%
\bibitem [{\citenamefont {Taghvaee}\ \emph {et~al.}(2020)\citenamefont
  {Taghvaee}, \citenamefont {Abadal}, \citenamefont {Pitilakis}, \citenamefont
  {Tsilipakos}, \citenamefont {Tasolamprou}, \citenamefont {Liaskos},
  \citenamefont {Kafesaki}, \citenamefont {Kantartzis}, \citenamefont
  {Cabellos-Aparicio},\ and\ \citenamefont {Alarcon}}]{Taghvaee2020}%
  \BibitemOpen
  \bibfield  {author} {\bibinfo {author} {\bibfnamefont {H.}~\bibnamefont
  {Taghvaee}}, \bibinfo {author} {\bibfnamefont {S.}~\bibnamefont {Abadal}},
  \bibinfo {author} {\bibfnamefont {A.}~\bibnamefont {Pitilakis}}, \bibinfo
  {author} {\bibfnamefont {O.}~\bibnamefont {Tsilipakos}}, \bibinfo {author}
  {\bibfnamefont {A.~C.}\ \bibnamefont {Tasolamprou}}, \bibinfo {author}
  {\bibfnamefont {C.}~\bibnamefont {Liaskos}}, \bibinfo {author} {\bibfnamefont
  {M.}~\bibnamefont {Kafesaki}}, \bibinfo {author} {\bibfnamefont {N.~V.}\
  \bibnamefont {Kantartzis}}, \bibinfo {author} {\bibfnamefont
  {A.}~\bibnamefont {Cabellos-Aparicio}},\ and\ \bibinfo {author}
  {\bibfnamefont {E.}~\bibnamefont {Alarcon}},\ }\bibfield  {title} {\bibinfo
  {title} {Scalability analysis of programmable metasurfaces for beam
  steering},\ }\href {https://doi.org/10.1109/access.2020.3000424} {\bibfield
  {journal} {\bibinfo  {journal} {{IEEE} Access}\ }\textbf {\bibinfo {volume}
  {8}},\ \bibinfo {pages} {105320} (\bibinfo {year} {2020})}\BibitemShut
  {NoStop}%
\end{thebibliography}%

\end{document}